\documentclass[aps,preprint]{revtex4}

\usepackage{amssymb}
\usepackage{amsmath}
\usepackage{epsfig}
\usepackage{epstopdf}
\usepackage{bm}
\usepackage{graphicx,epsfig}
\usepackage{mathrsfs}
\usepackage{dcolumn}
\usepackage{color}
\usepackage{natbib}
\usepackage{CJK}
\def\be{\begin{equation}}
\def\ee{\end{equation}}
\def\bea{\begin{eqnarray}}
\def\eea{\end{eqnarray}}

\allowdisplaybreaks[2]

\begin{document}

\title{Synthetic Dimension in Photonics}
\author{Luqi Yuan$^{1,4}$, Qian Lin$^2$, Meng Xiao$^{1}$, and Shanhui Fan$^{1,3}$}
\affiliation{$^1$Department of Electrical Engineering, and Ginzton
Laboratory, Stanford University, Stanford, CA 94305, USA }
\affiliation{$^2$Department of Applied Physics, Stanford
University, Stanford, CA 94305, USA}
\affiliation{$^3$E-mail:
shanhui@stanford.edu}
\affiliation{$^4$E-mail:
yuanluqi@stanford.edu}

\date{\today }

\begin{abstract}
The physics of a photonic structure is commonly described in terms
of its apparent geometric dimensionality. On the other hand, with
the concept of synthetic dimension, it is in fact possible to
explore physics in a space with a dimensionality that is higher as
compared to the apparent geometrical dimensionality of the
structures. In this review, we discuss the basic concepts of
synthetic dimension in photonics, and highlighting the various
approaches towards demonstrating such synthetic dimension for
fundamental physics and potential applications.
\end{abstract}


\maketitle

\section{Introduction}

The physics of a photonic structure is commonly described in terms
of its apparent geometric dimensionality.  A few 0, 1, 2 and 3
dimensional examples are microcavities \cite{ward11,feng12},
waveguides \cite{kawachi90}, two-dimensional photonic crystals
\cite{joannopoulos97}, and three-dimensional metamaterials
\cite{soukoulis11,liu11}, respectively.

Using these same photonic structures, however, it is in fact
possible to explore physics in a space with a dimensionality that
is higher as compared to the apparent geometrical dimensionality
of these structures. The basic idea is to configure
\textit{synthetic dimensions}, and to combine such synthetic
dimensions with the geometric dimensions to form higher
dimensional \textit{synthetic space}
\cite{regensburger11,regensburger12,schwartz13,luo15,yuanol,ozawa16}.

In photonics, there are two approaches for creating such a
synthetic space.

\begin{figure}[!h]
\centering
\includegraphics[height=0.45\linewidth]{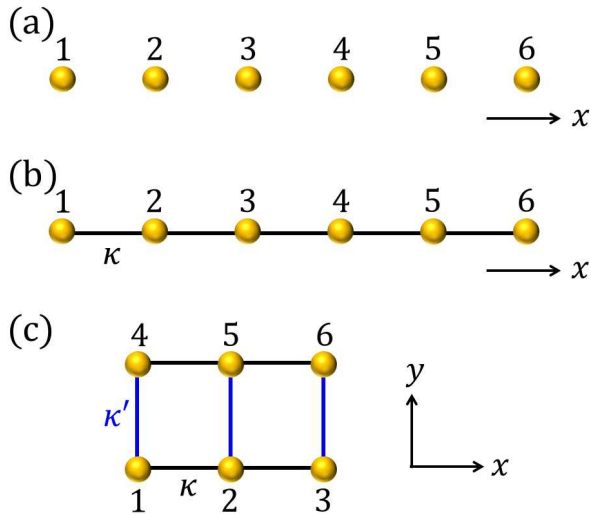}
\caption{(a) The physical states  labelled by consecutive
integers. (b) Introducing coupling between nearest-neighbor
physical states creates a one-dimensional system. (c) Introducing
 long-range coupling between different states generates a
two-dimensional system \cite{schwartz13,yuanhaldane}.
\label{Fig:connectivity}}
\end{figure}

\begin{enumerate}

\item Forming a lattice. Consider a system as described by
coupling a set of physical states together. The dimensionality of
the system is then determined by the nature of coupling. As an
illustration, suppose we label the states as consecutive integers,
and therefore we can visualize the states as being placed on a
line [Fig. \ref{Fig:connectivity}(a)]. The coupling in Fig.
\ref{Fig:connectivity}(b), where consists of a nearest neighbor
coupling between the states, results in a one-dimensional lattice.
On the other hand, with the same set of states, one can in fact
form higher-dimensional lattices, with the longer-range coupling
as indicated in Fig. \ref{Fig:connectivity}(c).

\item Exploiting the parameter dependency the system. As a general
illustration, consider a system in an \textit{n}-dimensional
space, as described by a Hamiltonian of the form
$H(p_1,\ldots,p_m,r_1,\ldots,r_n)$, where $p_1,\ldots,p_m$ are
external parameters, and $r_1,\ldots,r_n$ are the spatial
coordinates. One can think of each parameter of the Hamiltonian as
an extra synthetic dimension.

\end{enumerate}

In additional to photonics, both of these approaches for creating
synthetic dimensions have been also extensively explored in other
physical systems, such as cold atoms in optical lattices
\cite{boada12,kraus12,mei12,gomezleon13,ganeshan13,xu13,kraus13,celi14,mei14,boada15,price15,mancini15,stuhl15,nakajima16,lohse16,zeng16,suszalski16,taddia17,price17,martin17,baum18,peng18,lohse18,shang18}
and superconducting qubits \cite{strauch08,tsomokos10,mei16}. The
development of the concept of synthetic dimension in photonics
share some of the motivations. For examples, theoretically it is
known that there are rich physics, and in particular, rich
topological physics in systems beyond three dimensions
\cite{zhang01,qi08,kitaev09,lian16,lian17,roy17}. The development
of synthetic dimension provides an experimental approach to
explore such physics. Also, while three dimensional physics can in
principle be explored in three dimensional structures,
constructing such structures may be challenging and it may be more
advantageous to explore such a physics in one or two-dimensional
structures that are easier to construct.

On the other hand, there are aspects of synthetic dimension
concepts that are unique  in photonics. In particular, in forming
a lattice, the states that are being used can have different
frequencies, or different orbital angular momentums, corresponding
to different internal degrees of freedom of photons. The
construction of the synthetic space thus  enables new
possibilities for manipulating these internal degrees of freedom,
which are of significant potential importance for applications
such as communications and information processing.

In this paper, we provide a short review of the current
developments of the concepts of synthetic dimension in photonics.
The rest of the paper is organized as follows: In Sec. II, we
review the approach to create synthetic dimension by designing the
coupling between various photonic states to form a synthetic
lattice. In Section III, we discuss some of the physics effects in
these photonic synthetic lattices, focusing in particular on
photonic gauge potential and topological photonics effects. In
Sec. IV, we review the approach of exploring high dimensional
photonic phenomena in the parameter space. We conclude in Sec. V.

\section{Forming a synthetic lattice of photonic states}

In order to create a synthetic space using the approach of forming
a lattice, one needs a set of physical states, as well as
mechanisms to specifically configure the coupling among these
states. Photonics offers a rich variety of possibilities for
forming lattices. For the states, one can use photonic modes with
different frequencies, or  with different spatial distributions
such as different orbital angular momentums, or alternatively one
can use multiple temporal pulses. Photonic structures also offer
great flexibilities in configuring the coupling of these different
states. In this section, we brief review these various approaches
for creating synethic space by forming a lattice.

\subsection{Using photonic modes with different frequencies}

\begin{figure}[!h]
\centering
\includegraphics[height=0.67\linewidth]{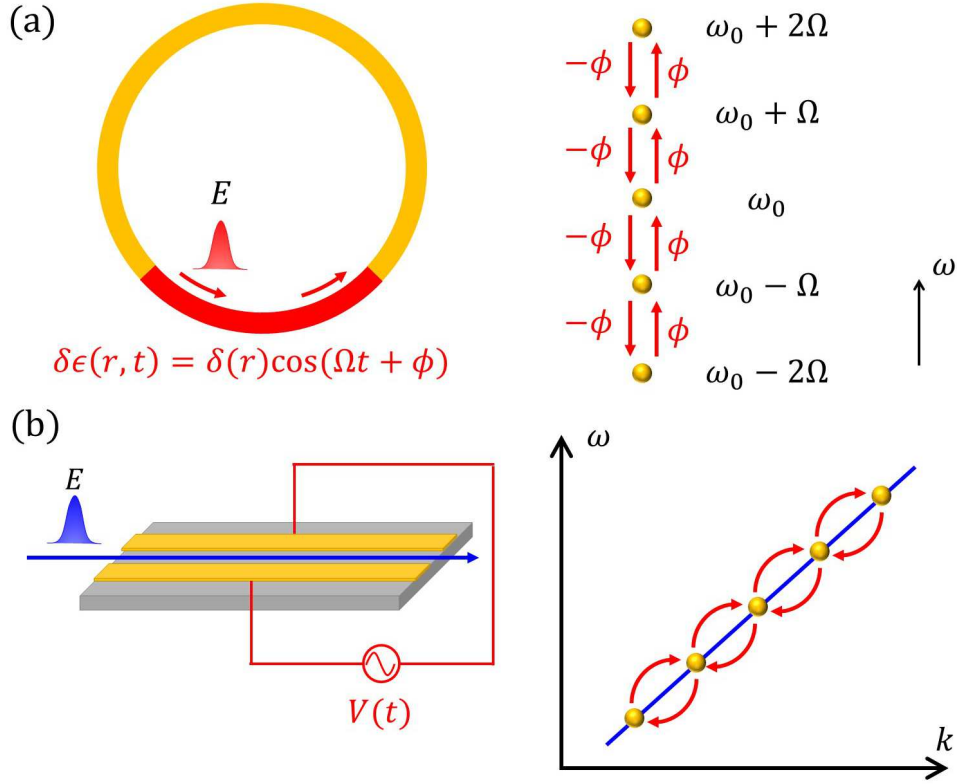}
\caption{(a) A dynamically modulated ring resonator can be
described by  a tight-binding model of a photon along a
one-dimensional lattice in the synthetic frequency dimension
\cite{yuanol,yuanoptica}. (b) A phase-matched modulation along a
dielectric waveguide can achieve a one-dimensional lattice formed
from waveguide modes at different frequencies \cite{qin18,bell17}.
\label{Fig:scheme}}
\end{figure}

In photonics, a natural way to create a synthetic dimension is to
use the frequency of light. Photonic structures naturally support
modes at different frequencies. Moreover, a lattice can be formed
by coupling these modes together, through either dynamic
modulation of the structure, or by nonlinear optics techniques.

As a simple illustration of the use of dynamic modulation to
couple modes with different frequencies together, consider a
structure with a time-dependent permittivity described by:
\begin{equation}
\epsilon(r,t)=\epsilon_s(r)+\delta \epsilon(r,t),
\label{Eq8:modulation}
\end{equation}
where $\epsilon_s(r)$ describes a static dielectric structure, and
$\delta \epsilon (r,t)$ describes a time-dependent modulation. For
such a time-dependent system, its electromagnetic properties can
be described from the Maxwell's equation as:
\begin{equation}
\nabla \times \nabla \times E + \mu_0 \frac{\partial ^2}{\partial
t^2} \epsilon_s E = - \mu_0 \frac{\partial ^2 P}{\partial t^2},
\label{Eq8:maxwell}
\end{equation}
where
\begin{equation}
P = \delta \epsilon E \label{Eq8:maxwellP}
\end{equation}
is the polarization current density induced by the dynamic
modulation.

Since for typical modulation, we have $|\delta \epsilon| \ll
\epsilon_s$, one can treat the dynamically modulated structure
perturbatively. We first determine the mode of the static
structure by solving an eigenvalue problem:
\begin{equation}
 \nabla \times \nabla \times E_m = \mu_0 \epsilon_s
\omega_m^2 E_m , \label{Eq8:maxwell2}
\end{equation}
where $\omega_m$ is the frequency of the mode, and $E_m$ is the
eigenmode field distribution. For the dynamic structure, the
modulation induces coupling between these modes. Therefore, we
expand the field $E$ in terms of the eigenmodes of the static
structure:
\begin{equation}
E = \sum_m a_m E_m e^{i\omega_m t}, \label{Eq8:maxwellE}
\end{equation}
and describe the properties of the dynamic structure in terms of
the dynamics of the modal amplitudes $a_m$$'s$.

The formalism above can be used to describe a modulated ring
resonator
\cite{yuanol,ozawa16,yuanoptica,linnc,zhang17,yuanhaldane,lin18,yuanPT}.
Consider a static ring resonator composed of a single-mode
waveguide, which we assume to have  zero group velocity dispersion
for simplicity. Suppose the ring supports a resonant mode at a
frequency $\omega_0$. In the vicinity of the frequency $\omega_0$,
resonant modes then form an equally spaced frequency comb,  with
the $m$-th resonant mode  having the frequency
\begin{equation}
\omega_m = \omega_0 + m\Omega_R. \label{Eq1:frequency}
\end{equation}
In Eq. (\ref{Eq1:frequency}), $\Omega_R = 2 \pi c/n_0 L$ is the
free-spectral range of the ring which also defines the modal
spacing in frequency, $n_0 = n(\omega_0)$ is the group index of
the waveguide at $\omega_0$, and $L$ is the circumference for the
ring.

Suppose we modulate the ring resonator structure above as:
\begin{equation}
\delta \epsilon(r,t) = \delta(r) \cos (\Omega t +\phi) ,
\label{Eq8:modulation}
\end{equation}
where $\delta(r)$ is the modulation profile, $\Omega$ is the
modulation frequency and $\phi$ is the modulation phase. For
simplicity we set $\Omega = \Omega_R$. The induced polarization
density from the $m$-th mode is then
\begin{equation}
P = \frac{\delta(r) E_m}{2}a_m \left( e^{i\phi} e^{i\omega_{m+1}t}
+ e^{-i\phi}e^{i\omega_{m-1}t} \right). \label{Eq8:maxwellP2}
\end{equation}
Thus, the induced polarization will resonantly excite modes $m+1$
and $m-1$. Hence we expect that the modal amplitudes satisfies:
\begin{equation}
i\frac{d a_m}{dt} = g e^{i\phi} a_{m-1} + g e^{-i\phi} a_{m+1}.
\label{Eq8:maxwellP2}
\end{equation}
Here $g$ is the coupling strength. (A detailed derivation can be
found in Refs. \cite{yuanol,ozawa16}.) The Hamiltonian of such a
system is then
\begin{equation}
H =g \sum_m \left(e^{i\phi} c^\dagger_{m+1} c_m + e^{-
i\phi}c^\dagger_m c_{m+1}\right). \label{Eq1:ham1ringRWA}
\end{equation}
where $c_m (c^\dagger_m)$ is the annihilation (creation) operator
for the $m$-th resonant mode. The Hamiltonian in Eq.
(\ref{Eq1:ham1ringRWA}) describes a tight-binding model of a
photon in a one-dimensional lattice in the synthetic frequency
dimension \cite{yuanol,yuanoptica}.

Similar approach for creating a synthetic dimension along the
frequency axis can also be achieved in a waveguide. Consider a
static single-mode waveguide with the propagation direction along
the $z$-direction, its eigenmode has the form
\begin{equation}
E_k = E_k (x,y) e^{-ikz}. \label{Eq8:waveguide}
\end{equation}
Here $E_k(x,y)$ is the modal profile, and $k$ is the wavevector.
The eigenfrequency of the mode is $\omega(k)$, which also defines
the dispersion relation of the waveguide. Suppose we operate in
the vicinity of an operating frequency $\omega_0$ with the
corresponding wavevector $k_0$, i.e. $\omega_0 = \omega(k_0)$.
Near $\omega_0$ one can expand the dispersion relation as:
\begin{equation}
\omega - \omega_0 = v_g (k-k_0), \label{Eq8:dispersion}
\end{equation}
where $v_g $ is the group velocity of the waveguide.

For such a waveguide, we then modulate its permittivity as:
\begin{equation}
\delta \epsilon = \delta (x,y) \cos (\Omega t - K z +\phi),
\label{Eq8:waveguidemodulation}
\end{equation}
where we choose the modulation to be phase-matched with the
waveguide mode, such that
\begin{equation}
\Omega = K v_g, \label{Eq8:waveguideK}
\end{equation}
In the presence of such modulation, we expand the field in the
waveguide as:
\begin{equation}
E = \sum_m a_m(z) E_m (x,y) e^{i\left(\omega_m t -k_m z\right)},
\label{Eq8:waveguideexpansion}
\end{equation}
where $\omega_m = \omega_0 + m\Omega$, and $k_m$ is the
corresponding wavevector, i.e. $\omega_m = \omega(k_m)$. The
induced polarization from the $m$-th mode has the form
\begin{equation}
P = \frac{\delta(x,y)E_m(x,y)}{2} a_m \left[e^{i\phi}
e^{i\left(\omega_{m+1}t-k_{m+1}z\right)} + e^{-i\phi}
e^{i\left(\omega_{m-1}t-k_{m-1}z\right)}\right],
\label{Eq8:waveguideP}
\end{equation}
We see that the induced polarization would couple with the $m+1$
and $m-1$ mode in a phase-matched fashion. We expect that the
coupled mode theory to have the form:
\begin{equation}
i \frac{d a_m }{dz} =  g e^{i\phi} a_{m-1} + g e^{-i\phi} a_{m+1}.
\label{Eq1:couplemodewaveguide}
\end{equation}
We again see a one-dimensional tight binding model for photons
along a synthetic frequency dimension.

The modulations in Eqs. (\ref{Eq8:modulation}) and
(\ref{Eq8:waveguidemodulation}) can be achieved using
electro-optical modulation \cite{qin18}. Recent developments of
on-chip silicon \cite{tzuang14} and LiNbO$_3$ modulators
\cite{wang14nb,zhang17nb} in either ring resonator or waveguide
geometries may prove to be quite useful for creating synthetic
lattice. In addition, one can also consider the use of
accoustic-optical modulators in fiber ring resonators. Similar
effects can also be accomplished with the use of nonlinear optical
effects \cite{peschel08,bersch09}.  Ref. \cite{bell17} used two
strong pumps differ in frequency by $\Omega$, to create a
synthetic lattice along the frequency dimension, for a weak probe
wave. Similar four-wave mixing process has also been considered in
a Raman medium \cite{yuanraman}.

\subsection{Using photonic modes with different orbital angular momentum}

\begin{figure}[!h]
\centering
\includegraphics[height=0.34\linewidth]{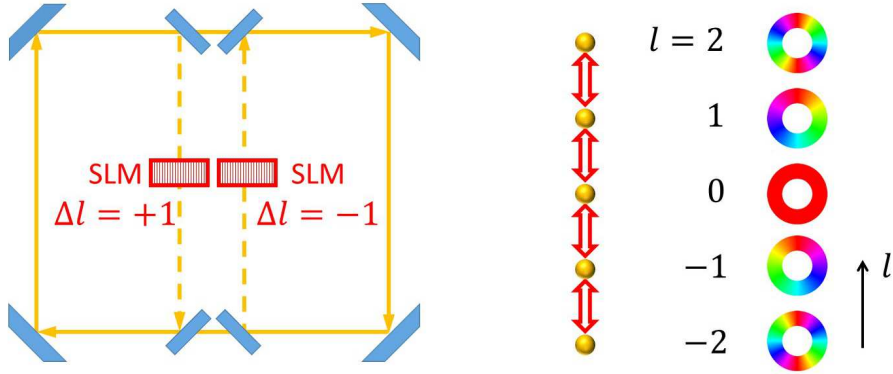}
\caption{A cavity that is degenerate for optical beams with
different angular momentum (solid line). Within this cavity, an
auxiliary cavity (dashed line) is incorporated, where two spatial
light modulators couple a beam at angular momentum $l$ to beams at
angular momentums $l \pm 1$, respectively. Such a cavity can be
described by a synthetic lattice along the angular momenta
direction \cite{luo15}. \label{Fig:scheme2}}
\end{figure}

Instead of creating a synthetic dimension in the spectral domain,
one can also create a synthetic dimension exploiting the spatial
degree of freedom in optical modes. Consider a light beam, with
its transverse profile carrying non-zero orbital angular momentum,
circulating around in a ring cavity to form resonant modes of the
cavity.  In general, the resonant frequency of such a resonant
mode should depend on the orbital angular momentum of the
corresponding circulating beam. However, with appropriate design,
one can in fact create a degenerate optical cavity, in which
resonant modes formed by beams with different orbital angular
momenta  have the same frequency. Inside such degenerate cavity,
one can then introduce an auxiliary cavity, which incorporates
spatial light modulators, in order to couple a small portion of
the amplitude of a beam with an orbital angular moment $l$  to a
beam with an orbital angular moment $l-1$ and $l+1$. The system
then is described by a tight-binding model
\cite{luo15,luo17,sun17,zhou17,luo18}
\begin{equation}
H= g \sum_l \left(e^{i\phi} c^\dagger_{l} c_{l-1}+e^{-i\phi}
c^\dagger_l c_{l+1}\right), \label{Eq1:hamOAM}
\end{equation}
where $l$ corresponds to the orbital angular moment of light.
Thus, the cavity structure shown in Fig. \ref{Fig:scheme2} can be
used to achieve a synthetic dimension based on the orbital angular
momentum. While here for simplicity we consider only nearest
neighbor coupling along the $l$-axis, long-range coupling can also
be achieved with different design of the spatial light modulators
(Ref. \cite{luo17}).

\subsection{Using multiple pulses}

\begin{figure}[!h]
\centering
\includegraphics[height=0.55\linewidth]{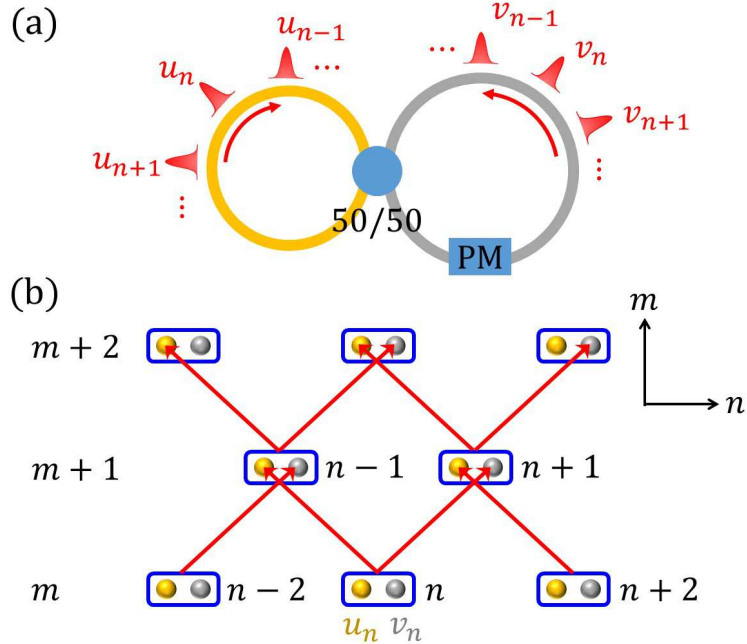}
\caption{(a) Two fibre loops connected by a $50/50$ coupler. (b)
An equivalent lattice network which describes a one-dimensional
synthetic lattice ($n$) evolves along the time ($m$)
\cite{regensburger11,regensburger12}. \label{Fig:pulses}}
\end{figure}

Another platform to create a synthetic photonic lattice is to
exploit the temporal degree of freedom, where the evolution of a
sequence of pulses is mapped onto the dynamics of a particle
moving on a set of discrete lattice sites
\cite{regensburger11,regensburger12,wimmer13,regensburger13,wimmer15,wimmer17,vatnik17,wimmer18}.
As an illustration, consider  two fibre loops with different
lengths,  connected by a $50/50$ coupler [see Fig.
\ref{Fig:pulses}(a)]. We assume that the round-trip times for
light travelling through the short and long loops are $T_s$ and
$T_l$, respectively, with the time difference $T_l-T_s \equiv
2\Delta T$ and the average time $(T_l+T_s) /2 \equiv T$. Within
the long loop there is in addition a phase modulator. For a pulse
at a particular position in  the short (long) loop at the time
$t_0$, after a round trip around the short (long) trip, it returns
to the same position at the time $t_0+T-\Delta T$ ($t_0+T+\Delta
T$). Suppose at the time $t=mT+n\Delta T$, two pulses, denoted as
$u_n^m$ and $v_n^m$, arrives at the input ends of the coupler in
the short and long loops, respectively. Two pulses $u^m_{n+1}$ and
$v^m_{n+1}$, upon passing through the coupler, generates an output
pulse in the short loop. Such an output pulse then goes through a
round trip in the short loop to generate $u^{m+1}_n$, i.e.
\begin{equation}
u_n^{m+1} = \frac{1}{\sqrt{2}} \left( u^m_{n+1} + i v^m_{n+1}
\right). \label{Eq1:hampulses1}
\end{equation}
And similarly
\begin{equation}
v_n^{m+1} = \frac{1}{\sqrt{2}} \left( i u^m_{n-1} + v^m_{n-1}
\right) e^{i\phi (n)}, \label{Eq1:hampulses2}
\end{equation}
In Eqs. (\ref{Eq1:hampulses1}) and (\ref{Eq1:hampulses2}), we have
used the scattering matrix of the coupler, and have incorporated
the effects of the phase modulators. Here we assume that the
modulation period is $T$, and hence the time-dependent
transmission phase depends on $n$ only. Eqs.
(\ref{Eq1:hampulses1}) and (\ref{Eq1:hampulses2}) describe the
temporal motion (motion along the $m$-axis) of a particle on a
one-dimensional synthetic lattice as labelled by $n$.

To summarize Section II, photonics provides a rich set of
opportunities to create synthetic lattices. The general idea here
is to  specifically design the coupling between various photonic
modes.  In addition to the few examples above, with different
spectral, spatial and temporal modes, there exist many other
possibilities of utilizing photonic modes. For example, Ref.
\cite{lustig18} experimentally demonstrated a synthetic lattice
that maps into a topological insulator, based on modes in an array
of coupled waveguides. While in the examples above for simplicity
we have considered one-dimensional synthetic lattices with nearest
neighbor coupling, it is possible to achieve higher dimensional
synthetic lattice with more complex couplings. For example, Refs.
\cite{schwartz13,yuanhaldane} show that by utilizing modulators
with a few modulation frequencies, one can achieve a synthetic
lattice with dimensions higher than one using the same set of
modes in the ring resonator here as we considered in Section II.
A. The effects of long-range coupling in synthetic dimensions have
been also explored in Ref. \cite{bell17}.

In photonics, the number of distinct lattice sites along the
synthetic dimension can potentially be quite large. For ring
resonators systems, for example, it is conceivable to have
hundreds of different modes coupling together, since the
modulation frequency is typically far smaller than the resonant
frequencies of the modes.  Have such a large space along the
synthetic dimension is useful for the demonstration of analogues
of bulk physics effects in the synthetic space. In addition, the
boundaries along the synthetic dimension can also be introduced,
either naturally through the group velocity dispersion for the
modulated ring or waveguides \cite{yuanol}, or by specifically
designed boundaries using a memory effect \cite{baum18}.

\section{The Physics of Synthetic Lattice}

The photonic synthetic lattices as discussed in Section II provide
versatile platforms to explore fundamental physics effects. In
addition, since the synthetic lattice is built upon various
degrees of freedom of light, the abilities to control the flow of
light in the synthetic lattice provides abilities to control
properties of light that are important for practical applications.

The description of a dynamically modulated ring in terms of a
one-dimensional tight-binding model certainly has a long history.
This description, for example, has been used to described the
physics of mode-locked lasers
\cite{harris65,harris67,haus75,haus00}. In recent years, the
concept of the synthetic lattice has been explored to demonstrate
a wide range of physics effects including the physics of
parity-time symmetry
\cite{regensburger12,regensburger13,wimmer15,yuanPT}, Anderson
localization \cite{vatnik17}, and time-reversal of light
\cite{wimmer18}. Here we focus on two important emerging
directions in the physics of synthetic lattice: creating an
effective gauge potential for light, and topological photonics
effects.

\subsection{Effective gauge potential}

Photons are neutral particles. Thus, there is no naturally
occurring gauge potential that couples to photons. On the other
hand, in the construction of photonic synthetic lattices, the
ability for achieving an effective gauge potential for photons
naturally emerge. To illustrate the concept of such an effective
gauge potential for photons \cite{fang12,fang12np,yuanprl},
consider first the Hamiltonian
\begin{equation}
H= g \sum_{\langle i,j\rangle} \left( e^{-i\phi_{ij}(t)}
c_i^\dagger c_j + e^{i\phi_{ij} (t)} c_j^\dagger c_i\right),
\label{Eq2:ham}
\end{equation}
where $\phi_{ij}(t)$ is the hopping phase between lattice sites
$i$ and $j$. The hopping phase in general can be time-dependent.
For simplicity we consider only nearest neighbor coupling.  We can
therefore make the association \cite{luttinger51}
\begin{equation}
\int_i^j A \cdot d\mathbf{r} = \phi_{ij}(t). \label{Eq2:gauge}
\end{equation}
where $A$ is the effective gauge potential for photons. In a
higher dimensional lattice \cite{fang12np},
\begin{equation}
B = \frac{1}{S} \oint _\mathrm{plaquette} A \cdot d\mathbf{r}
\label{Eq2:magnetic}
\end{equation}
is the effective magnetic field through a plaquette, here $S$ is
the area of the plaquette. Also, if $\phi_{ij}(t)$ is time
dependent, we then have a time dependent gauge potential $A(t)$
\cite{yuanprl}.
\begin{equation}
E = - \frac{\partial A}{\partial t} \label{Eq2:electric}
\end{equation}
is therefore an effective electric field for photons. As we see in
Section II, the various techniques for creating a photonic
synthetic lattice naturally incorporates the capabilities for
controlling such hopping phases in the lattice. Therefore, these
technique naturally leads to gauge potentials, as well as
effective electric or magnetic fields for photons.

As perhaps the simplest illustration of the effect of a gauge
potential, we consider the effect of Bloch oscillation in the
synthetic space along the frequency axis. Bloch oscillation occurs
when a charge particle in a one-dimensional lattice is subject to
a constant electric field. This effect, initially proposed in
solid state physics \cite{bloch29}, has been previously considered
for photons in waveguide arrays
\cite{lenz99,trompeter06,longhi06,dreisow09,levy10} and photonic
crystals
\cite{kavokin00,malpuech01,sapienza03,agarwal04,lousse05,estevez15}.
Here we show that such an effect can occur in the synthetic space
as well \cite{yuanoptica,longhi05,bersch09,peschel08,bersch11}.
Consider the ring resonator incorporating a phase modulator as
discussed in Section II. A. By choosing the modulation frequency
$\Omega$ to be slightly different from the mode spacing
$\Omega_R$, the resulting Hamiltonian has the same form as Eq.
(\ref{Eq1:ham1ringRWA}), but with a time-dependent phase $\phi(t)
= (\Omega-\Omega_R) t$. From Eq. (\ref{Eq2:electric}), then, such
a modulation results in an effective time-independent electric
field in the synthetic space.

\begin{figure}[!h]
\centering
\includegraphics[height=0.5\linewidth]{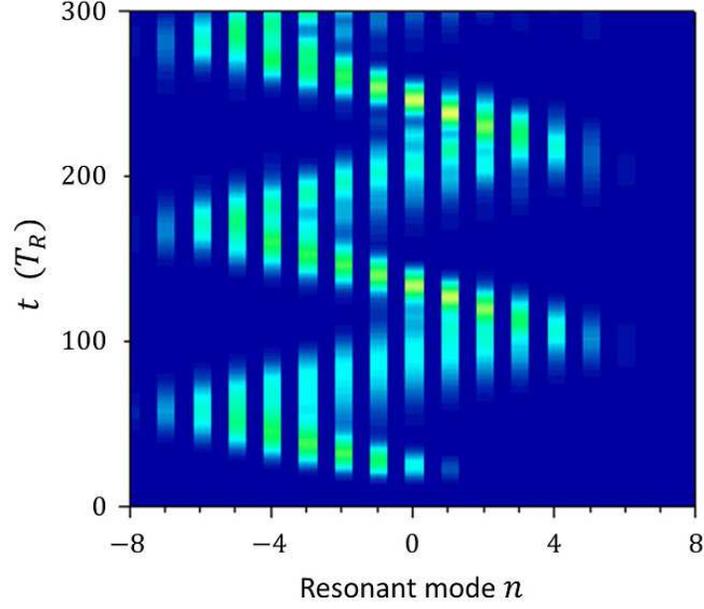}
\caption{The evolution of the light in time for each resonant mode
$n$ exhibits the spectral Bloch oscillation \cite{yuanoptica}.
\label{Fig:physicsbloch}}
\end{figure}

The effect of such a constant effective electric field can be seen
in Fig. \ref{Fig:physicsbloch}. Suppose at $t=0$ a few spectral
components are excited. As time evolves the excited spectral
components oscillates, which is precisely the effect of Bloch
oscillation  in the spectral domain. Moreover, it was noted in
Ref. \cite{yuanoptica} that a periodic switching of the modulation
frequency around the mode spacing can give rise to  a
uni-directional shift of photons along the frequency axis, which
is a useful capability for controlling the frequency of light.
Related capabilities for controlling the spectrum of light,
including negative refraction and focusing of light along the
frequency axis, has also been demonstrated by creating a photonic
gauge potential in a waveguide
\cite{bell17,peschel08,bersch09,qin18}, based on the waveguide
system as discussed in Section II. A.

\begin{figure}[!h]
\centering
\includegraphics[height=0.5\linewidth]{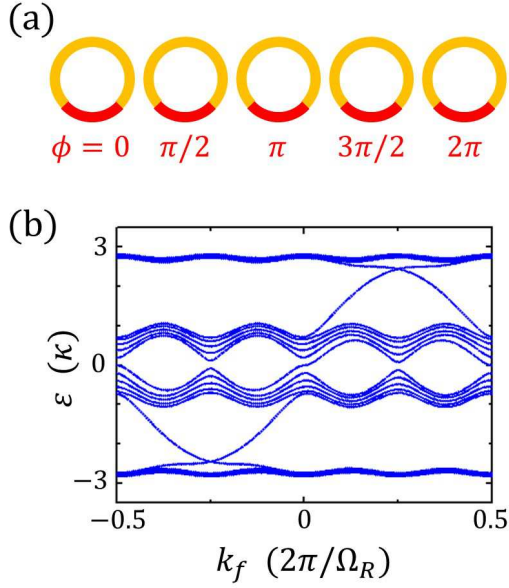}
\caption{(a) A one-dimensional array of ring resonators undergoing
dynamic modulations \cite{yuanol,ozawa16}. The modulators in the
rings have different modulation phases $\phi$.  (b) Projected band
structure for the system in (a) with 21 rings, assuming that the
lattice is infinite along the synthetic frequency dimension
\cite{yuanol}. \label{Fig:physicsgauge}}
\end{figure}

One can also explore the consequences of effective magnetic fields
for photons in the synthetic space
\cite{yuanol,ozawa16,yuanraman}. For this purpose, consider an
array of identical ring resonators, each of which incorporates a
phase modulator with the modulation frequency equal to the mode
spacing in the ring [Fig. \ref{Fig:physicsgauge}(a)]. Based on the
discussion in Section II. A, each ring supports a one-dimensional
lattice along the frequency axis. The modes in nearest-neighbor
rings at the same frequency  are also coupled through evanescent
tunnelling. The system thus is described by a Hamiltonian:
\begin{equation}
H =  \sum_{m,n} g \left(e^{- in\phi}c^\dagger_{m,n} c_{m+1,n} +
e^{i n\phi} c^\dagger_{m+1,n} c_{m,n}\right) + \kappa
\left(c^\dagger_{m,n} c_{m,n+1} + c^\dagger_{m,n+1}
c_{m,n}\right), \label{Eq1:hamringarray}
\end{equation}
where $m$ denotes the different modes in the same ring, $n$ labels
a ring in the array, and $\kappa$ is the coupling constant due to
evanescent tunnelling between two nearest-neighbor rings. We
assume that the spacing between the rings is $d$. By choosing the
modulation phase in Eq. (\ref{Eq8:modulation}) to be $n\phi$ for
the modulator in the $n$-th ring, as shown in Eq.
(\ref{Eq1:hamringarray}), the resulting Hamiltonian gives rise to
a uniform effective magnetic field $\phi/\Omega_R d$ in the
synthetic space as described in the Landau gauge.

The Hamiltonian of Eq. (\ref{Eq1:hamringarray}) is periodic along
the frequency axis (i.e. $m$-axis). Thus the wavevector reciprocal
to the frequency axis, $k_f \in [-\pi/\Omega_R,\pi/\Omega_R)$, is
conserved. Such $k_f$ conservation remains true even with a finite
number of rings. In Fig. \ref{Fig:physicsgauge}(b), we plot the
projected bandstructure, which represents the eigenvalues of the
Hamiltonian in Eq. (\ref{Eq1:hamringarray}) as a function of
$k_f$, for a system consisting of  21 rings with the choice of
$\phi =\pi/2$ \cite{yuanol}. The bandstructure exhibits one-way
edge states, as expected for such a lattice system in the presence
of an effective magnetic field. This system therefore enables
one-way frequency translation that is topologically protected.
Hamiltonian similar to Eq. (\ref{Eq1:hamringarray}) can also be
created in the synthetic space based on orbital angular momentum
of light as discussed in Section II. B \cite{luo15}. In such a
case the system provides a novel platform for controlling and
converting the orbital angular moment of light, which is of
potential importance for communication applications.

\subsection{Topological Photonics}

In the previous section, the Hamiltonian of Eq.
(\ref{Eq1:hamringarray}) in fact is topologically non-trivial. The
resulting band structure has non-trivial Chern number that arises
from the effective magnetic field. In addition to using such
magnetic field, there are other mechanisms in the synthetic
systems to achieve non-trivial topological effects. For example,
in a system similar to what has been discussed in Section II. B,
where modes with different orbital angular momenta are coupled
together to form a one-dimensional synthetic lattice, one can
realize the Su-Schriffer-Heeger (SSH) model with a sharp boundary,
and hence demonstrating bulk-edge correspondence in the SSH model
\cite{zhou17}.

\begin{figure}[!h]
\centering
\includegraphics[height=0.55\linewidth]{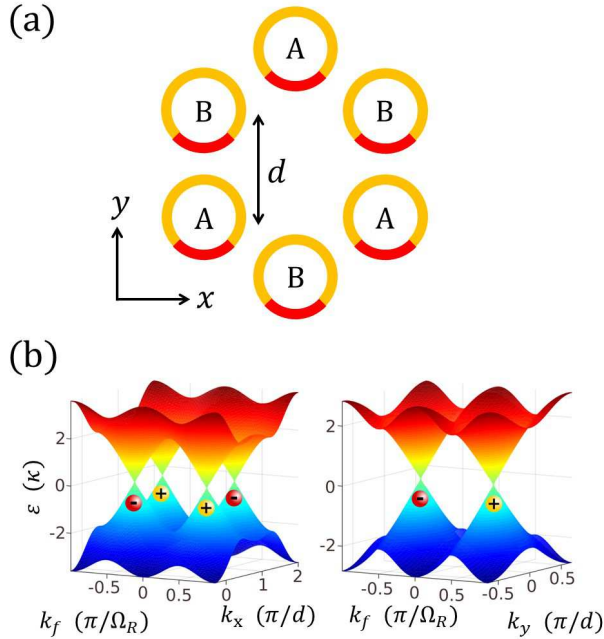}
\caption{(a) A two-dimensional honeycomb array of ring resonators
undergoing dynamic modulations. (b) Bandstructures show Weyl
points in three-dimensional synthetic space. Left (right): the
band structure in $k_x (k_y) - k_f$ plane at $k_y=0$ ($k_x =
4\pi/3\sqrt{3}d$) \cite{linnc}. \label{Fig:physicsweyl}}
\end{figure}

The effects of topological physics depends strongly on the
dimensionality of the physical system. There are rich set of
effects that are unique to higher-dimensional system with no
lower-dimensional counterparts. The concept of synthetic dimension
provides a natural pathway towards exploring these
higher-dimensional physics.

Here we illustrate this pathway by considering the exploration of
Weyl point physics in synthetic dimension
\cite{linnc,zhang17,sun17}.  A Weyl point is a two-fold degeneracy
in a three-dimensional band structure with linear dispersion in
its vicinity  \cite{weyl29}. The Weyl point  is of substantial
interest in topological physics because it represents a magnetic
monopole in the momentum space, and hence its presence is robust
to any small perturbation.

Photonic structures that support a Weyl point typically have
complex three-dimensional geometries
\cite{bravoabad15,lu15,chen15,wang16weyl}. On the other hand, with
the concept of synthetic dimension, it is in fact possible to
explore Weyl-point physics with two-dimensional geometries  that
are easier to construct. Consider a two-dimensional array of ring
resonators forming a honeycomb lattice. Each ring resonator
incorporates a phase modulator.  Therefore, as discussed in
Section II. A, each ring supports a one-dimensional lattice along
the frequency axis. The system thus is described by a
three-dimensional lattice model. It was shown in Ref. \cite{linnc}
that Weyl-point physics emerges by appropriately choosing the
modulation phases $\phi_A$ and $\phi_B$, on the $A$ and $B$ sites
of the honeycomb lattices (Fig. \ref{Fig:physicsweyl}).  Along
similar directions, Ref. \cite{lin18} showed that a
three-dimensional topological insulator can be constructed using a
two-dimensional ring resonator lattice. Four-dimensional quantum
Hall effect has been studied in a three-dimensional resonator
lattice \cite{ozawa16}. Also, it was shown that the
two-dimensional Haldane model can be implemented using only three
ring resonators \cite{yuanhaldane}.

In Section II we have discussed various techniques for achieving
long-range coupling in the synthetic space. Such long range
coupling can be used to create topological flat bands
\cite{sun11,tang11} which is important for simulating of many-body
physics including the fractional quantum Hall effect
\cite{umucalilar12}. The presence of long-range coupling can also
be used for achieving novel band structure effects such as a
single Dirac cone in a two-dimensional system without breaking
time-reversal symmetry \cite{mross16}.

Exploration of nonlinear effects in synthetic space is certainly
of fundamental interests in the context, for example, of quantum
simulations. In many important interacting lattice Hamiltonians,
the interaction is local with respect to the lattice sites. On the
other hand, for the schemes involving the frequency axis as the
synthetic dimension, typical nonlinear optics effects result in a
form of interaction that is nonlocal across different lattice
sites \cite{strekalov16}, and thus is not directly suitable for
the simulation of local-interacting Hamiltonians. Ref.
\cite{ozawa17} proposed to achieve local interaction in a system
where the synthetic dimension is the geometrical angular
coordinate. It is an interesting open question to achieve local
interaction for other approaches aiming to create synthetic space.

\section{Synthetic dimension from  parameter space}

\subsection{Physics concept}

Instead of forming a synthetic lattice,  another common method to
create a synthetic space is to utilize the parameter degrees of
freedom. Consider any physical system as described by a
Hamiltonian $H(p)$ that is parametrically dependent on a
continuous variable $p$.  The parameter dependency of the system
can be alternatively described in a synthetic space with the
$p$-axis as an extra synthetic dimension in addition to the usual
physical space. In this way, higher dimensional physics can then
manifest in terms of parameter dependency of a lower-dimensional
physical system.

The concept of gauge potential and the associated topological
physics effects naturally arise in such synthetic space
incorporating the parameter axis. As  a simple illustration,
consider a Hamiltonian in the parameter space described by a
two-dimensional vector $\mathbf{R}$, which satisfies the
Schr\"{o}dinger equation \cite{berry84,raffaele00}
\begin{equation}
H (\mathbf{R}) |\Psi (\mathbf{R})\rangle = E (\mathbf{R}) |\Psi
(\mathbf{R})\rangle , \label{Eq2:hampara}
\end{equation}
We assume that as $\mathbf{R}$ varies the Hilbert space does not
change. Suppose we consider a closed curve $\mathcal{C}$ in the
$R$-space. Along this closed loop, one can define the Berry's
phase $\gamma$ as
\begin{equation}
\gamma = \oint_\mathcal{C} i \langle \Psi (\mathbf{R})
|\nabla_{\mathbf{R}}| \Psi(\mathbf{R})\rangle \cdot d\mathbf{R}.
\label{Eq2:berryphase}
\end{equation}
The integral kernel here gives the Berry connection, or the gauge
potential in parameter space:
\begin{equation}
\mathbf{A} (\mathbf{R})=  i \langle \Psi (\mathbf{R})
|\nabla_{\mathbf{R}}| \Psi(\mathbf{R})\rangle .
\label{Eq2:berryconnection}
\end{equation}
With the Stokes's theorem, we obtain the Berry curvature
\begin{equation}
\mathbf{B}(\mathbf{R}) =  \nabla_{\mathbf{R}} \times
\mathbf{A}(\mathbf{R}). \label{Eq2:berryflux}
\end{equation}
The usual topological description of a two-dimensional band
structure can be formulated in exactly the same fashion as above,
with $\mathbf{R}$ corresponding to the wavevector $k$
\cite{roushan14,schroer14,yale16}. Since the wavevector is defined
on the first Brillouin zone which is a 2-torus, the integration of
the $B$-field is quantized and gives rise to the Chern numbers of
the bands. The important observation here, however, is that the
topological argument commonly used for band-structure can in fact
be used for any parameter dependency. Moreover, if one were to
vary the parameter as a function of time adiabatically, the
dynamics of such time-dependent system has signature of higher
dimensional topological physics.  Below, we illustrate these
aspects with specific examples.

\subsection{Nontrivial topology in the parameter space}

\begin{figure}[!h]
\centering
\includegraphics[height=0.65\linewidth]{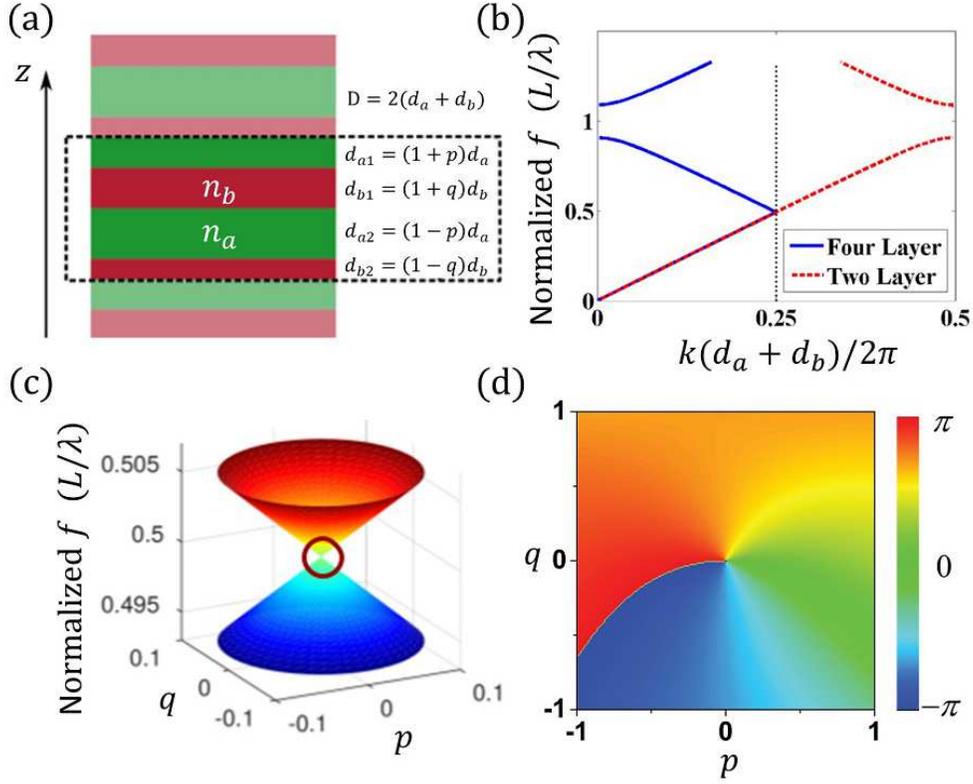}
\caption{(a) A one-dimensional photonic crystal in Ref.
\cite{wang17} with each unit cell including four layers where the
thickness of each layer depends on parameters $p$ and $q$. (b) The
band structure of the crystal with $p=q=0$. (c) Bandstructure in
the ($p$, $q$) space with $k=\pi/2(d_a+d_b)$. (d) The reflection
phase for light incident upon such a photonic crystal from air, in
the ($p$, $q$) space \cite{wang17}. \label{Fig:parameter}}
\end{figure}

As an illustration, consider a one-dimensional photonic crystal in
Ref. \cite{wang17}, the unit cell of which consists of four
layers, with thickness $(1+p)d_a$, $(1+q)d_b$, $(1-p)d_a$, and
$(1-q)d_b$. The band structure of such a crystal, in the special
case where $p=q=0$, is shown in Fig. \ref{Fig:parameter}(b). In
such a case, the four-layer unit cell is in fact not a primitive
cell, and thus there is no band gap at the edge of the first
Brilliouin zone $k=\pi/2(d_a+d_b)$. At this $k$ point the bands
are two-fold degenerate with linear dispersion along the $k$-axis.
On the other hand, for $p$ and $q$ slightly deviating from 0, the
four-layer unit cell becomes the primitive cell, and hence a band
gap opens at the Brillouin zone edge, as shown in Fig.
\ref{Fig:parameter}(c). The size of the gap scales linearly with
respect to both $p$ and $q$. Therefore, one can show that in the
three dimensional space of $k$, $p$ and $q$, the point
$k=\pi/2(d_a+d_b)$, $p=0$, and $q=0$ is in fact a Weyl point. The
physical signature of such a Weyl point manifests in the
reflection phase for a wave at a frequency of the Weyl point
$f=0.49$ $L/\lambda$, incident from air onto the photonic crystal
along the normal incidence direction. As one varies the parameters
$p$ and $q$, the reflection phase winds around $p=q=0$.
Remarkably, in this construction one can explore aspects of
three-dimension Weyl-point physics  using a simple one-dimensional
structure.

\subsection{Adiabatic evolution in the parameter space}

\begin{figure}[!h]
\centering
\includegraphics[height=0.6\linewidth]{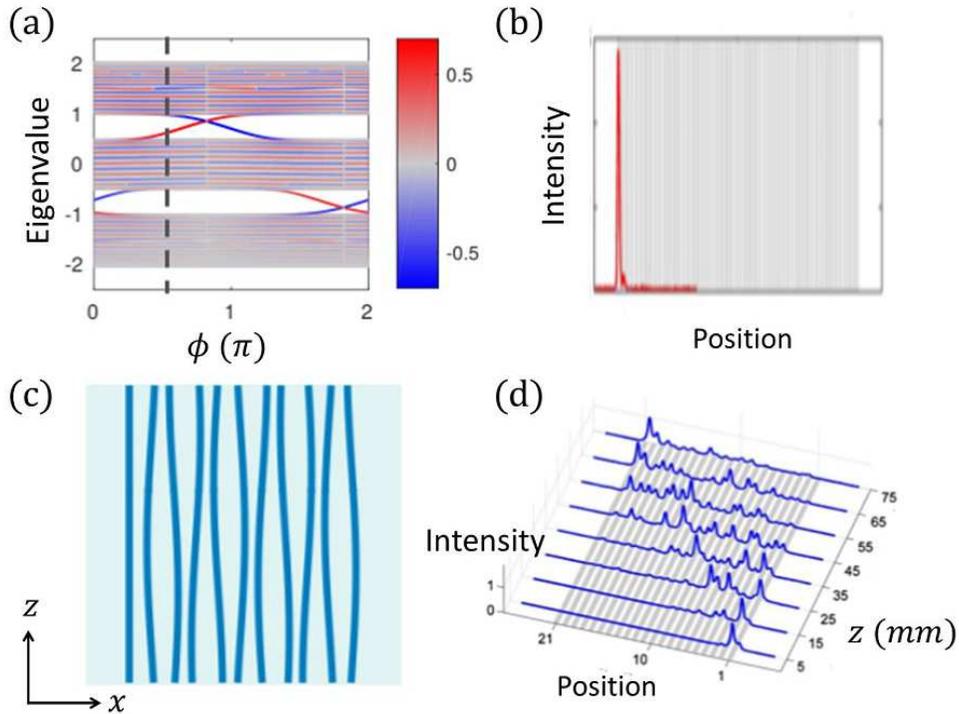}
\caption{(a) Bandstructure of the Aubry-Andr\'{e} model  described
by Eq. (\ref{Eq1:hamquasi}) composed of 99 sites with $g=1$,
$V=0.5$, and $b=(\sqrt{5}+1)/2$. (b) Intensity distributions in
the experiment show the edge state in the gap. Here $\phi=0.5\pi$.
(c) Waveguide array where spacings between waveguides are slowly
modified along the $z$-direction, as  described by Eq.
(\ref{Eq1:hamquasi2}). (d) Intensity distributions versus the
propagation distance $z$. Light is injected into the rightmost
waveguide into a structure similar to (c). $\phi$ is changed from
$0.35\pi$ to $1.75\pi$ adiabatically \cite{kraus12quasi}.
\label{Fig:quasicrystal}}
\end{figure}

In the previous section, for a system as described by a parametric
Hamiltonian, the higher-dimensional physics is revealed by
considering the properties of a set of physical structures with
varying parameters. On the other hand, with a single physical
structure, one can also explore higher dimensional physics by
allowing the parameters to vary as a function of time, and
considering the dynamics of such time-dependent parametric system.

As an illustration we consider consider the Aubry-Andr\'{e} model
\cite{aubry80} which describes a one-dimensional lattice
\begin{equation}
H = \sum_m g \left( a^\dagger_m a_{m+1} + a^\dagger_{m} a_{m-1}
\right) + \sum_m a^\dagger_m a_m V \cos(2\pi b m +\phi),
\label{Eq1:hamquasi}
\end{equation}
where $a_m (a^\dagger_m)$ is the annihilation (creation) operator
on the $m$-th site, and $g$ is the coupling strength. $V$ is the
amplitude of the on-site potential. $b$ and $\phi$ are parameters
controlling the modulation of the on-site potential with respect
to the site locations \cite{kraus12quasi}. We note that for an
irrational $b$ there is no periodicity in Eq.
(\ref{Eq1:hamquasi}). In fact, in that case Eq.
(\ref{Eq1:hamquasi}) describes a quasi-crystal.

By considering $\phi$ as a synthetic dimension, the
Aubry-Andr\'{e} model describes a synthetic two-dimensional space.
In fact, it is known that Aubry-Andr\'{e} model is closely related
to the Quantum Hall system as described by a two dimensional model
of a square lattice under a uniform magnetic field in the Landau
gauge, as described by the Hamiltonian of Eq.
(\ref{Eq1:hamringarray}). The $b$ in Eq. (\ref{Eq1:hamquasi}) maps
to the magnetic field per unit cell in the two-dimensional model,
and  $\phi$ maps to the wavevector $k_y$ along periodic direction
with the choice of the Landau gauge. Thus, for a finite structure,
the eigenspectrum of the Aubry-Andr\'{e} model exhibits gaps [Fig.
\ref{Fig:quasicrystal}(a)]. For particular values of $\phi$, an
edge state can appear inside the gap [Fig.
\ref{Fig:quasicrystal}(b)]. The variation of eigenfrequencies as a
function of $\phi$ corresponds directly to the dispersion of the
one-way edge state in the Quantum Hall system.

In the finite-size Aubry-Andr\'{e} model, suppose at $t = 0$ the
system is at an edge state on one of ends, by varying $\phi$
adiabatically as a function of time, the mode will evolve
according to the edge state dispersion, eventually becomes a bulk
state, and then reemerges as an edge state localized on the
opposite end. Thus, the adiabatic evolution of the state in the
time-dependent one-dimensional system provides a direct probe of
the properties of the corresponding two-dimensional system.

Ref. \cite{kraus12quasi} provides a direct experimental
demonstration of such adiabatic evolution of  the edge state. For
this purpose, Ref. \cite{kraus12quasi} made two important
modification of the Aubry-Andr\'{e} model. First of all, the
Aubry-Andr\'{e} model is transformed to:
\begin{equation}
H = g \sum_m \left[1+V \cos(2\pi b m +\phi)\right] a^\dagger_m
a_{m+1} + \left[1-V \cos(2\pi b (m-1) +\phi)\right] a^\dagger_{m}
a_{m-1}. \label{Eq1:hamquasi2}
\end{equation}
where the site-dependent modulation now appears in the coupling
constant between nearest neighbor sites. Secondly,  instead of
considering time evolution, Ref. \cite{kraus12quasi} consider an
array of waveguides, where the variation of the field amplitudes
along the propagation direction $z$ then provides a simulation of
the temporal dynamics. With these modifications, Ref.
\cite{kraus12quasi} constructed a structure as shown in Fig.
\ref{Fig:quasicrystal}(c), where $\phi$ varies from $0.35\pi$ to
$1.75\pi$ as a function $z$. The adiabatic dynamics is shown in
\ref{Fig:quasicrystal}(d). The injected light into the edge state
at one end of the waeguide array evolves into a bulk state as
light propagates along the $z$-direction, and eventually appear as
an edge state on the other side, which is precisely the adiabatic
evolution of the state as expected for the Aubry-Andr\'{e} model.

The use of adiabatic evolution provides a powerful approach to
explore higher dimension physics. In the waveguide array platform,
this approach has also been used to experimentally explore
topological phase transition \cite{verbin13}, as well as
four-dimensional quantum Hall effects \cite{zilberberg18}.  The
connection of the one-dimensional Aubry-Andr\'{e} model, which can
be quasiperodic with an irrational $b$, to a two dimensional model
also in fact points to a general connection between quasi-crystal
in lower dimensional space and crystal structure in higher
dimensional space. This connection has been previously explored to
develop a computational tool for photonic quasicrystal
\cite{rodriguez08}.

\section{Summary and outlook}

To summarize this article, we provide a brief review of the
concept of synthetic dimension in photonics, an area that has been
rapidly developing in the past few years, in close connection with
the developments of the gauge field and topological concepts in
photonics. The initial motivation for exploring the synthetic
dimension has been to develop a versatile approach in photonics
for demonstrating many important fundamental physics effects,
including in particular many important topological physics
effects. And indeed, as we have seen in this review, a remarkably
rich set of physics effects have been theoretically proposed
and/or experimentally demonstrated  using the synthetic dimension
approach. We also envision that the concept of synthetic dimension
will prove to be significant for practical applications, leading
to new possibilities for manipulating and controlling some of the
fundamental properties of light.

\begin{acknowledgments}
This work is supported by a Vannevar Bush Faculty Fellowship from
the U. S. Department of Defense (Grant No. N00014-17-1-3030), the
U. S. Air Force Office of Scientific Research Grant No.
FA9550-17-1-0002), and the National Science Foundation (Grant No.
CBET-1641069).
\end{acknowledgments}

\end{document}